\documentclass[aps,prc,twocolumn,floatfix,nofootinbib]{revtex4}
\usepackage{graphicx}
\usepackage[usenames,dvipsnames]{color}
\usepackage{amsmath,amssymb,bbold}
\usepackage{hyperref}
\usepackage{comment}

\begin{document}

\title{Correspondence between Israel-Stewart and first-order casual and stable hydrodynamics for the boost-invariant massive case with zero baryon density}

\author{Arpan Das$^{1}$}
\email{arpan.das@ifj.edu.pl}
\author{Wojciech Florkowski$^{2}$}
\email{wojciech.florkowski@uj.edu.pl}
\author{Radoslaw Ryblewski${^1}$}
\email{radoslaw.ryblewski@ifj.edu.pl}

\affiliation{$^{1}$Institute  of  Nuclear  Physics  Polish  Academy  of  Sciences,  PL-31-342  Krakow,  Poland}

\affiliation{$^{2}$Institute of Theoretical Physics, Jagiellonian University, PL-30-348 Krakow, Poland}

\begin{abstract}
Exact correspondence between Israel-Stewart theory and first-order causal and stable hydrodynamics is established for the boost-invariant massive case with zero baryon density and the same constant relaxation times used in the shear and bulk sectors. Explicit expressions for the temperature dependent regulators are given for the case of a relativistic massive gas. The stability and causality conditions known in the first-order approach are applied and one finds that one of them is violated in this case. 
\end{abstract}

\maketitle
\section{Introduction}

The success of relativistic hydrodynamics as a main tool used for modeling heavy-ion collisions~\cite{Florkowski:2010zz,Gale:2013da,Jeon:2015dfa, Jaiswal:2016hex,Busza:2018rrf} triggered broad interest in general aspects of this theory~\cite{Romatschke:2017ejr,Florkowski:2017olj}. It has turned out that the formalism of relativistic hydrodynamics includes many interesting new features such as the asymptotic character of the hydrodynamic gradient expansion \cite{Heller:2013fn,Denicol:2016bjh,Heller:2016rtz,Grozdanov:2019kge} or the existence of hydrodynamic attractors~\cite{Heller:2015dha,Romatschke:2017vte,Strickland:2017kux,Strickland:2018ayk,Jaiswal:2019cju,Giacalone:2019ldn}. 

The phenomenological hydrodynamic models used to analyze the data are based on the Israel-Stewart (IS) version of this theory \cite{Israel:1976tn,Israel:1979wp} --- as already in the 1970's it was realized that the formulations derived earlier by Landau and Eckart were not causal~\cite{Hiscock:1983zz,Hiscock:1985zz}. The IS approach treats the shear stress tensor $\pi^{\mu\nu}$ and the bulk pressure $\Pi$ as new hydrodynamic variables, in addition to temperature $T$ and hydrodynamic flow $u^\mu$ (in a baryon free case). In most of the studied cases, the IS theory is stable and causal, which is essential for its practical applications. In the meantime, certain disadvantages of the IS formulation have been also removed (see, for example, Ref.~\cite{Florkowski:2010cf,Martinez:2010sc,Denicol:2012cn,Jaiswal:2013npa,Jaiswal:2013vta,Tinti:2015xwa,Attems:2018gou,Montenegro:2018bcf,Calzetta:2019dfr}).

Very recently, a completely new hydrodynamic approach has been proposed by F.~S.~Bemfica, M.~M.~Disconzi, J.~Noronha, and P.~Kovtun~\cite{Bemfica:2017wps,Bemfica:2019knx,Kovtun:2019hdm}. It treats $T$ and $u^\mu$ as fundamental hydrodynamic variables and is based on a first-order expansion in derivatives. It also employs the opportunity of a more general choice of the hydrodynamic frame and introduces a new set of kinetic coefficients that play the role of ultra-violet regulators of the theory. They make it causal (even in the full nonlinear regime) and linearly stable around equilibrium. Below we refer to this theory with the acronym FOCS, as it is first-order, causal, and stable (if the proper choice of the regulators is made). 

A natural question can be asked about possible relations between the IS and FOCS formulations. In general no direct connection between these two theories exists, as IS leads to ten differential equations, while FOCS gives four second-order equations which are equivalent to only eight equations of the first order. Nevertheless, there may exist special cases where the two frameworks lead to the same dynamical equations. Such cases are interesting and useful as they allow us to ``transfer'' the knowledge gained in one sector to the other one. In particular, the information about causality and stability established for the FOCS approach can be used to analyze IS solutions, provided such connections exist. 

In our previous paper \cite{DAS2020135525} we have found that there is an exact matching between FOCS and IS for boost-invariant, Bjorken expanding systems with a massless, conformal equation of state, $p = \frac{1}{3} \varepsilon$, and a regulating sector determined by a constant relaxation time $\tau_R$. We have also studied there the stability and causality properties of this model. In this work, we extend these investigations to the case of systems of massive particles. 

As we have noticed above, no direct connections between FOCS and IS are expected, as they lead to a different number of equations and unknown functions. As a matter of fact, in the massive and boost-invariant case FOCS yields two equations, while IS leads to three equations (for $T$, one independent component of the shear stress tensor which we denote here as $\pi$, and the bulk pressure $\Pi$). In this work, we demonstrate that there is a special choice of the IS framework and its kinetic parameters such that the three IS equations contain the two FOCS boost-invariant equations. As in the massless case, the matching found between the two frameworks can be used to learn more about the two theories by transferring the knowledge established for one formulation to the other one (and vice versa). In this work we use the causality and stability criteria established for FOCS to check the properties of the corresponding IS framework.

The paper is organized as follows: In the next section we introduce the IS and FOCS hydrodynamic equations and discuss the formula for the bulk viscosity coefficient~$\zeta$. In Sec. III we construct a matching between the two frameworks.  The case of massive particles obeying classical statistics is discussed in more detail in Sec.~IV. In Sec.~V we apply the FOCS causality and stability criteria. We summarize and conclude in Sec. VI. Throughout the paper we use natural units $\hbar = c = k_B =1$.

\section{IS and FOCS frameworks}  

\subsection{Israel-Stewart boost-invariant set-up}
 
In this work we follow the IS boost-invariant version of hydrodynamic equations for baryon-free systems defined by Eqs.~(23)--(25) in Ref.~\cite{Jaiswal:2013fc}. They read as follows:
 \begin{align}
  & \frac{d\varepsilon}{d\tau} = -\frac{1}{\tau} \bigg[(\varepsilon+p)+\Pi-\pi\bigg],\label{equ8}\\
  & \tau_{\pi}\frac{d\pi}{d\tau}= \frac{4}{3}\frac{\eta}{\tau} -\pi -\beta \frac{\tau_{\pi}}{\tau}\pi,
  \label{equ9}\\
  & \tau_{\Pi}\frac{d\Pi}{d\tau}= -\frac{\zeta}{\tau} -\Pi -\beta \frac{\tau_{\Pi}}{\tau}\Pi.
  \label{equ10}
 \end{align}
Here $\varepsilon, p, \pi$ and $\Pi$ are the energy density, pressure, shear stress, and bulk pressure, respectively. They are all functions of the longitudinal proper time $\tau = \sqrt{t^2-z^2}$. The parameters $\tau_\pi$ and $\tau_\Pi$ are relaxation times, while $\eta$ and $\zeta$ are the shear and bulk viscosity coefficients. In Eqs.~\eqref{equ9} and \eqref{equ10} the original value of $\beta$ is $4/3$, however, below we treat $\beta$ as a free parameter (with exception of Sec.~V). We note that Eqs.~\eqref{equ8}--\eqref{equ10} represent one of the simplest versions of the IS hydrodynamic framework --- we do not consider richer structures since for the massive systems they are not very helpful to establish the IS--FOCS correspondence. 

Following Ref.~\cite{DAS2020135525} we introduce the time derivative of temperature as a separate variable, $y = dT/d\tau$, and rewrite Eq.~\eqref{equ8} as
\begin{align}
\pi -  \Pi = \frac{d\varepsilon}{dT} \,y \,\tau + (\varepsilon+p),
 \label{equ11}
\end{align}
which after differentiation with respect to $\tau$ yields
\begin{align}
 \dot{\pi}-\dot{\Pi} = \frac{d^2\varepsilon}{d T^2}y^2\tau+\frac{d\varepsilon}{d T}\dot{y}\tau+\left(\frac{d p}{d T}+2\frac{d \varepsilon}{\partial T}\right)y.
 \label{equ14}
\end{align}
Here the dot denotes the proper time derivative, i.e., $\dot{T}\equiv dT/d\tau$.

Since the difference $\pi\!-\!\Pi$ appears in Eq.~\eqref{equ8}, it is useful to construct the second equation containing the same difference from a linear combination of Eqs.~(\ref{equ9}) and (\ref{equ10}). If the new equation does not include the term $\pi\!+\!\Pi$, we obtain a system of two coupled equations that may match the FOCS boost-invariant formulation. In this case, the remaining equation constructed from Eqs.~\eqref{equ9}--\eqref{equ10}, which includes the combination $\pi\!+\!\Pi$ only, remains decoupled from the first two equations. 

It is easy to check that the situation described above happens if the relaxation times in the shear and bulk sectors are the same. Consequently, in what follows we assume that $\tau_{\pi}=\tau_{\Pi}=\tau_R$, as in \cite{Jaiswal:2013fc}, and that $\tau_R$ is constant, as in \cite{DAS2020135525}. Using these assumptions, from  Eqs.~\eqref{equ9} and \eqref{equ10} we obtain
\begin{align}
 \tau_R (\dot{\pi}-\dot{\Pi}) 
 & = \left(\frac{4}{3}\frac{\eta}{\tau}+\frac{\zeta}{\tau}\right)-\left(1+\beta\frac{\tau_R}{\tau}\right)(\pi-\Pi).
 \label{equ12}
\end{align}

Substituting Eq.~\eqref{equ11} into the right-hand side of Eq.~\eqref{equ12} and Eq.~\eqref{equ14} into the left-hand side we obtain
\begin{align}
 & \tau_R\frac{d\varepsilon}{d T}\dot{y}+\tau_R \frac{d^2\varepsilon}{d T^2}y^2+y\bigg[\frac{\tau_R}{\tau}\left(\frac{d p}{d T}+2\frac{d\varepsilon}{d T}\right)\nonumber\\
 & +\left(1+\beta\frac{\tau_R}{\tau}\right)\frac{d\varepsilon}{d T}\bigg]+\bigg[\left(1+\beta\frac{\tau_R}{\tau}\right)\frac{\varepsilon+p}{\tau}\nonumber\\
 &~~~~~~~~~~~~~~~~~~~~~~~~~~~~~-\left(\frac{4}{3}\frac{\eta}{\tau^2}+\frac{\zeta}{\tau^2}\right)\bigg] = 0.
 \label{equ15}
\end{align}
In the last expression one can recognize the Ricatti equation studied earlier in~\cite{Denicol:2017lxn}.

\subsection{FOCS boost-invariant set-up}

For FOCS one uses the constitutive relations in the form~\cite{Bemfica:2019knx}
\begin{align}
 \mathcal{E}(\tau)=\varepsilon(\tau)+\varepsilon_1\frac{\dot{T}}{T}+\frac{\varepsilon_2}{\tau},
 \label{equ16}\\
 \mathcal{P}(\tau)=p(\tau)+\pi_1\frac{\dot{T}}{T}+\frac{\pi_2}{\tau}
 \label{equ17}
\end{align}  
or equivalently as
\begin{align}
    & \mathcal{E}= \varepsilon +\chi_1 \left(\frac{d\varepsilon}{d T}\right)\frac{T}{\varepsilon+p}\frac{\dot{T}}{T}+\frac{\chi_2}{\tau},\nonumber\\
    & \mathcal{P}=p+ \chi_3 \left(\frac{d\varepsilon}{d T}\right)\frac{T}{\varepsilon+p}\frac{\dot{T}}{T}+\frac{\chi_4}{\tau}.\nonumber
\end{align}
Hence, we can make the following identifications between the regulators $\varepsilon_i$, $\pi_i$ and $\chi_i$:
\begin{eqnarray}
\varepsilon_1 &=& \chi_1\frac{T}{\varepsilon+p}\left(\frac{d\varepsilon}{d T}\right), \quad \varepsilon_2 = \chi_2, 
\label{eq:eps1ch1} \\
\pi_1 &=& \chi_3\frac{T}{\varepsilon+p}\left(\frac{d\varepsilon}{d T}\right), \quad  \pi_2 = \chi_4.
\label{eq:eps2ch2}
\end{eqnarray}
The bulk viscosity in FOCS is expressed in terms of the $\chi_i$ coefficients~\cite{Bemfica:2019knx}
\begin{align}
\zeta = \chi_3-\chi_4+c_s^2(\chi_2-\chi_1),
\end{align}
where $c_s^2 = dp/d\varepsilon$ is the sound velocity squared. Using Eqs.~(\ref{eq:eps1ch1}) and (\ref{eq:eps2ch2}), and the thermodynamic relations \mbox{$\varepsilon + p = Ts$} and $dp = s dT$, we find
\begin{align}
\zeta = c_s^2\pi_1-\pi_2+c_s^2
\left(\varepsilon_2-c_s^2\varepsilon_1\right).
\label{equ26}
\end{align}

\medskip
Using Eqs.~\eqref{equ16} and \eqref{equ17} we find
\begin{align}
 \frac{d \mathcal{E}}{d \tau}
& = \frac{d\varepsilon}{d T}\dot{T}+\frac{d\varepsilon_1}{d T}\frac{1}{T}\dot{T}^2 + \frac{\varepsilon_1}{T}\ddot{T} \nonumber\\
&~~~~~~~~
-\frac{\varepsilon_1}{T^2}\dot{T}^2
+\frac{1}{\tau}\frac{d\varepsilon_2}{d T}\dot{T}-\frac{\varepsilon_2}{\tau^2}
\label{equ18}
 \end{align}
and
\begin{align}
 \frac{\mathcal{E}+\mathcal{P}}{\tau} = \frac{\varepsilon+p}{\tau}+\frac{\varepsilon_1}{\tau}\frac{\dot{T}}{T}+\frac{\pi_1}{\tau}\frac{\dot{T}}{T}+\frac{\varepsilon_2}{\tau^2}+\frac{\pi_2}{\tau^2}.
 \label{equ19}
\end{align}
Equations~\eqref{equ18} and \eqref{equ19} allow us to write the boost-invariant FOCS equation 
\begin{align}
 & \frac{d\mathcal{E}}{d\tau}+\frac{\mathcal{E}+\mathcal{P}}{\tau}-\frac{4}{3}\frac{\eta}{\tau^2} = 0
\end{align}
as
\begin{align}
& \frac{\varepsilon_1}{T}\dot{y} +y^2\bigg(\frac{d \varepsilon_1}{d T}\frac{1}{T}-\frac{\varepsilon_1}{T^2}\bigg)\nonumber\\
& +y\bigg[\frac{d\varepsilon}{d T}+\frac{1}{\tau}\frac{d \varepsilon_2}{d T}+\frac{\varepsilon_1}{\tau}\frac{1}{T}+\frac{\pi_1}{\tau}\frac{1}{T}\bigg]\nonumber\\
& +\bigg[\frac{\varepsilon+p}{\tau}+\frac{\pi_2}{\tau^2}-\frac{4}{3}\frac{\eta}{\tau^2}\bigg] = 0,
\label{equ20}
\end{align}
where again we use the notation $y = \dot{T}$ (it should be interpreted as the second differential equation).

\section{Matching the two approaches}

It is easy to notice that Eq.~\eqref{equ20}, similarly to Eq.~\eqref{equ15}, has the form of the Ricatti equation. Hence, the IS and FOCS frameworks may become equivalent if the coefficients appearing in Eqs.~\eqref{equ15} and~\eqref{equ20} can be made equal. Comparing the coefficients multiplying the time derivative $\dot{y}$ we find
\begin{align}
 \varepsilon_1 = \tau_R \frac{d\varepsilon}{d T}T.
 \label{eq:eps1}
\end{align}
Comparing the coefficients standing at $y^2$ in Eqs.~\eqref{equ15} and~\eqref{equ20} we find that
\begin{align}
 \tau_R \frac{d^2\varepsilon}{d T^2} = \frac{d\varepsilon_1}{d T}\frac{1}{T}-\frac{\varepsilon_1}{T^2}.
 \label{equ22}
\end{align}
It can be easily shown that $\varepsilon_1$ given by Eq.~\eqref{eq:eps1} also satisfies Eq.~\eqref{equ22}, hence, Eq.~\eqref{equ22} becomes irrelevant. 

In the next step it is convinient to compare the terms in Eqs.~\eqref{equ15} and Eq.\eqref{equ20} that do not contain $y$. This leads to the relation
\begin{align}
\left(1\!+\!\beta\frac{\tau_R}{\tau}\right)\frac{\varepsilon\!+\!p}{\tau}-\frac{4}{3}\frac{\eta}{\tau^2}-\frac{\zeta}{\tau^2}=\frac{\varepsilon\!+\!p}{\tau}+\frac{\pi_2}{\tau^2}-\frac{4}{3}\frac{\eta}{\tau^2}.
\end{align}
The equation above can be used to determine $\pi_2$, namely
\begin{align}
\pi_2 = \beta \tau_R (\varepsilon + p)-\zeta.
\label{eq:pi2}
\end{align}
Using Eq.~\eqref{equ26} in Eq.~\eqref{eq:pi2} we obtain a formula connecting $\pi_1$, $\varepsilon_1$, and $\varepsilon_2$,
\begin{align}
\pi_1=\frac{\beta\tau_R}{c_s^2}(\varepsilon+p)-\left(\varepsilon_2-c_s^2\varepsilon_1\right).
\label{eq:pi1}
\end{align}
We recall that Eq.~\eqref{eq:eps1} defines the coefficient $\varepsilon_1$ in terms of the IS parameters. So if we knew $\varepsilon_2$, then using \eqref{eq:pi1} we could determine $\pi_1$. The missing constraint comes from the comparison of the terms multiplying $y$ in  Eqs.~\eqref{equ15} and~\eqref{equ20}, which gives
\begin{eqnarray}
&& \frac{d\varepsilon}{d T}+\frac{1}{\tau}\frac{d\varepsilon_2}{d T}+\frac{\varepsilon_1}{\tau T}+\frac{\pi_1}{\tau T}  \nonumber \\
&& \,\,\, =  \frac{\tau_R}{\tau}\left(\frac{d p}{d T}+2 \frac{d \varepsilon}{d T}\right)  +\left(1+\beta\frac{\tau_R}{\tau}\right)\frac{d \varepsilon}{d T}.
\label{eq:yconstraint}
\end{eqnarray}
With the help of the thermodynamic identities listed above,  Eqs.~\eqref{eq:pi1} and \eqref{eq:yconstraint} lead to the differential equation for the function $\varepsilon_2(T)$,
\begin{eqnarray}
\frac{d\varepsilon_2}{d T}-\frac{\varepsilon_2}{T} = \tau_R \frac{d\varepsilon}{d T}.
\label{eq:eps2eq}
\end{eqnarray}
A formal solution of Eq.~\eqref{eq:eps2eq} is
\begin{eqnarray}
    \varepsilon_2(T) = \tau_R \, T \!\! \int_{T_0}^T \!\! \frac{\,\,\,d\varepsilon(T')}{T' d T'}~dT',
    \label{eq:eps2sol}
\end{eqnarray}
where $T_0$ is an integration constant that we set equal to zero.

Consequently, if the equation of state is known, i.e., the temperature dependence of the energy density, $\varepsilon = \varepsilon(T)$, is given then from Eq.~\eqref{eq:eps2sol} we find $\varepsilon_2(T)$. Once $\varepsilon_2$ is known, from Eq.~\eqref{eq:pi1} we can find $\pi_1$. The remaining two coefficients, $\varepsilon_1$ and $\pi_2$ come from Eqs.~\eqref{eq:eps1} and \eqref{eq:pi2}, respectively. Hence, all the coefficients appearing in FOCS can be uniquely detrmined in terms of the IS coefficients $\eta$ and $\zeta$, as well as the equation of state.

Interestingly, even if the equation of state is not known, we can find direct relations connecting different regulators.  Since $d\varepsilon = T d s$, Eqs.~\eqref{eq:eps1} and \eqref{eq:eps2sol} give:
\begin{eqnarray}
\varepsilon_1 =  \,c_s^{-2}\, \tau_R T  s, \quad
\varepsilon_2 = \tau_R  T  s.
\label{eq:eps2sol2}
\end{eqnarray}
Hence, the second expression in the round brackets on the right-hand side of Eq.~\eqref{eq:pi1} vanishes and we find
\begin{align}
\pi_1= \beta c_s^{-2} \,  \tau_R(\varepsilon+p) = 
\beta  c_s^{-2} \tau_R T s  = \beta \varepsilon_1.
\label{eq:pi12}
\end{align}
Similarly, using \eqref{eq:pi2} one finds
\begin{align}
\pi_2 = \beta \tau_R T s - \zeta = \beta \varepsilon_2 - \zeta,
\label{eq:pi22}
\end{align}
thus, as long as $\zeta \ll \beta \varepsilon_2$ we expect $\pi_2 \approx \beta \varepsilon_2$. Expressing the $\chi_i$ coefficients by $\varepsilon_i$ and $\pi_i$, we also find that
\begin{equation}
\chi_1=\chi_2=\frac{\chi_3}{\beta}=\frac{\chi_4+\zeta}{\beta} =\tau_R T s >0.
\end{equation}

\section{Relativistic gas}

In this section, to illustrate our procedure, we take into consideration the equation of state of a relativistic gas. We first analyze the massless case and subsequently turn to a discussion of the massive gas obeying classical statistics. 

\subsection{Massless limit}

As a limiting case of our procedure, we analyze massless particles with a conformal equation of state $\varepsilon=3p=aT^4$, vanishing bulk viscosity $\zeta = 0$, and $c_s^2=1/3$. The parameter $a$ is a constant proportional to the number of internal degrees of freedom of particles. The FOCS regulators are obtained from  Eqs.~\eqref{eq:eps1}, \eqref{eq:pi2}, \eqref{eq:pi1} and~\eqref{eq:eps2sol}:
\begin{eqnarray} 
\varepsilon_1^{(m=0)} &=& 4a\tau_RT^4 ,
\nonumber \\
 \varepsilon_2^{(m=0)} &=& \frac{4}{3}a\tau_R  T^4,
 \nonumber \\
 \pi_1^{(m=0)} &=& 
 \beta \varepsilon_1^{(m=0)}, 
  \nonumber \\
 \pi_2^{(m=0)} &=& 
 \beta \varepsilon_2^{(m=0)}. 
 \label{eq:all0}
\end{eqnarray}
These expressions agree with our previous results obtained with $\lambda=0~$\cite{DAS2020135525}.

\begin{figure}[t]
\begin{center}
\includegraphics[width=9.5cm]{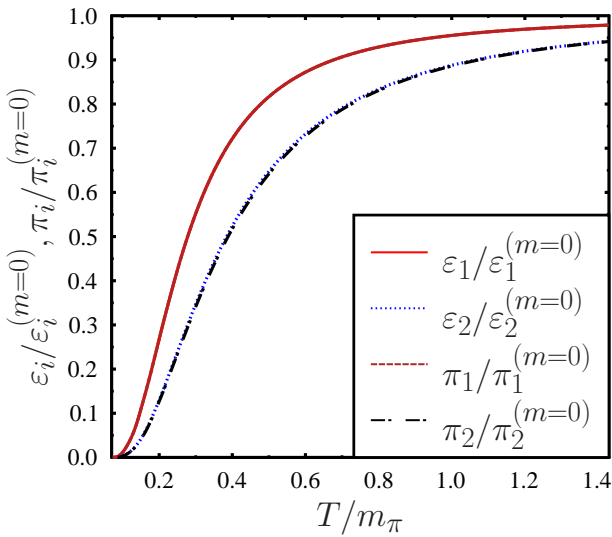}
\end{center}
\caption{\small Temperature dependence of the regulators $\varepsilon_1$, $\varepsilon_2$, $\pi_1$, and $\pi_2$ obtained for the equation of state given by Eqs.~\eqref{eq:varepsm} with $m = m_\pi =$ 140 MeV. The results are scaled by the values corresponding to the $m~\to~0$ limit Eqs.~\eqref{eq:all0}. }
\label{fig:massive case}
\end{figure}

\subsection{Massive classical case}

Let us now generalize our results to the massive, classical case. In this case the energy density and pressure are given by the modified Bessel functions $K_n$,
\begin{eqnarray}
\varepsilon(T) &=& \frac{a}{6} \, T^4  z^2 \bigg[ 3 K_2\left(z \right) + z K_1\left(z \right) \bigg],\nonumber\\
p(T) &=& \frac{a}{6} \, T^4  z^2   K_2\left(z \right),
\label{eq:varepsm}
\end{eqnarray}
respectively, where $z=m/T$ and we set $a=3g/\pi^2$ (with $g$ being the degeneracy factor) in order to agree with previous expressions given in the massless limit. 

The results of our numerical calculations for the regulators $\varepsilon_1$, $\varepsilon_2$, $\pi_1$, and $\pi_2$, where we used the value of the pion mass for $m$, are shown in Fig.~\ref{fig:massive case}. In the calculation of $\pi_2$ we used the formula for the bulk viscosity coefficient given in~\cite{Florkowski:2014sfa},
\begin{eqnarray}
\!\!\!\zeta\!=\!\tau_R \frac{p \, z^{2}}{3}\left[-\frac{z K_{2} }{3\left(3K_{3}  +z K_{2} \right)}\!+\!\frac{z}{3}\left(\frac{K_{1}}{K_{2}}-\frac{K_{i, 1}}{K_{2}}\right)\right],
\label{eq:oldzeta}
\end{eqnarray}
where all Bessel functions are understood to be evaluated at $z$ and $K_{i,1}(z)\!=\!\frac{\pi}{2} \left[1 - z K_0(z) L_{-1}(z) - z K_1(z) L_0(z) \right]$ with $L_i$ being the modified Struve function. The result \eqref{eq:oldzeta} has been obtained within the relaxation-time approximation (RTA) in the kinetic theory \cite{ANDERSON1974466}. We have identified here the RTA and IS relaxation times. The quantity $p$ in  \eqref{eq:oldzeta} is the same equilibrium pressure as that appearing in hydrodynamic equations. 

As expected the numerical calculations confirm that $\pi_1 = \beta \varepsilon_1$. We also find that to a very good approximation $\pi_2 \approx \beta \varepsilon_2$, which indicates that the considered values of the bulk viscosity are relatively small. In the high temperature limit, the values of the regulators approach their massless limits given by Eqs.~\eqref{eq:all0}.

Using Eq.~\eqref{eq:eps1} and expanding it around $z=0$ one finds deviations from the high-temperature limit, namely
\begin{eqnarray}
\varepsilon_{1} = \varepsilon_{1}^{(m=0)}  \left[1-\frac{z^2}{24}-\frac{z^4}{192}+{\cal O}(z^6)   \right].
\end{eqnarray}
The expansion of $\pi_1$ confirms that $\pi_1 = \beta \varepsilon_1$. On the other hand, using Eq.~\eqref{eq:eps2sol} one finds
\begin{eqnarray}
\varepsilon_{2} = \varepsilon_{2}^{(m=0)} \left[1-\frac{z^2}{8}+\frac{z^4}{64}+{\cal O}(z^6)   \right]
\end{eqnarray}
and
\begin{eqnarray}
\pi_{2} = \pi_{2}^{(m=0)} \left[1-\frac{z^2}{8}+\frac{z^4}{64}
\left(1\!-\!\frac{20}{27\beta}\right)+{\cal O}(z^6)\right],
\end{eqnarray}
so the scaling $\pi_2 = \beta \varepsilon_2$ is only approximate.

An analogous analysis at large $z$ gives
\begin{eqnarray}
\varepsilon_1=\kappa z  \!\left[ 128 z^2 + 624 z + 1785 + {\cal O}\left(\frac{1}{z}\right)\right]
\end{eqnarray}
and $\pi_1 = \beta \varepsilon_1$, where for simplicity of notation we have defined the function
\begin{eqnarray}\kappa (T,z) = \frac{1}{3072} \sqrt{\frac{\pi }{2}} \, \varepsilon_1^{(m=0)}  e^{-z} \sqrt{z}.
\end{eqnarray}
One similarly obtains
\begin{eqnarray}
\varepsilon_2=\kappa \left[ 128 z^2 + 560 z +945+ {\cal O}\left(\frac{1}{z}\right)\right]
\end{eqnarray}
and
\begin{eqnarray}
\pi_2&=&\kappa \beta \left[ 128   z^2 + 560z \left(1+\frac{16}{105\beta}\right) \right. \nonumber\\
&+& \left. 945 \left(1 + \frac{928}{2835\beta}\right)+ {\cal O}\left(\frac{1}{z}\right)\right].
\end{eqnarray}
This leads to an approximate relation $\pi_2 \approx \beta \varepsilon_2$ if the terms containing inverse powers of $\beta$ can be neglected.

\section{Stability and causality conditions}

Finally, we turn to the discussion of causality and stability conditions derived for FOCS in \cite{Bemfica:2019knx}. For the reader's convenience and also in order to be concrete we list them below and divide into two classes:

\smallskip
\begin{widetext}
\noindent  Causality conditions:
\begin{align}
& \hspace{2cm}\lambda_{\rm {\smallskip BDN}}>0, 
\quad \chi_1>0, \quad \eta\ge 0,  \quad  \lambda_{\text{BDN}}\ge \eta, 
\quad \beta\ge \frac{\phi}{\chi_1},  \label{eq:causal1}  \\
&\hspace{4cm} \chi_1\bigg[\lambda_{\text{BDN}}(c_s^2 + \beta ) + \phi\bigg] \ge 0, \\
&
\hspace{2cm}\lambda_{\text{BDN}}\chi_1\bigg[1- c_s^2  \left(\frac{\phi}{\chi_1}-\beta\right)\bigg]\ge \chi_1\bigg[\lambda_{\text{BDN}}(c_s^2 + \beta ) + \phi\bigg],\\
&9 \chi_1^2\bigg\{\lambda_{\text{BDN}}^2   c_s^4+2 \lambda_{\text{BDN}}  c_s^2 \left[\lambda_{\text{BDN}}\left(\frac{2\phi}{\chi_1}-\beta \right)+\phi\right]+ \bigg[ \phi+\lambda_{\text{BDN}} \beta\bigg]^2 \bigg\}\ge0.  
\label{equstability3}
\end{align}

\smallskip
\noindent Stability conditions:
\begin{eqnarray}
 && \hspace{6cm} \phi \ge 0, \label{eq:zetaeta} \\ 
&& 9  \phi \left\{c_s^2 \left(\lambda_{\text{BDN}}+\chi_1\right)\left(\lambda_{\text{BDN}}^2+\lambda_{\text{BDN}}\chi_1+\chi_1^2\right) +
  \chi_1 \bigg[\chi _1 \phi +\lambda_{\text{BDN}} \beta (\lambda_{\text{BDN}}+\chi_1)\bigg]\right\}\ge 0.\\
 && \hspace{4.5cm} (\lambda_{\text{BDN}} +\chi _1)(1-c_s^2) \ge \phi.
\label{eq:complcond}
\end{eqnarray}
\end{widetext}

The above conditions, where $\phi=4 \eta/3 +\zeta$, have been checked by us for the case of the relativistic massive gas discussed in the previous section. Besides the bulk viscosity coefficient, for which we have used Eq.~\eqref{eq:oldzeta}, the causality and stability conditions involve the shear viscosity coefficient $\eta$. We have used the formula for $\eta$ from Ref.~\cite{Florkowski:2014sfa} that reads
\begin{eqnarray}
\eta = \frac{\tau_R \,p\, z^3}{15} \left[
\frac{3}{z^2} \frac{K_3}{K_2} -\frac{1}{z} +
\frac{K_1}{K_2} - \frac{K_{i,1}}{K_2} \right].
\label{eq:oldeta}
\end{eqnarray}
The remaining unknown function appearing in the conditions \eqref{eq:causal1}--\eqref{eq:complcond} is the energy flow coefficient $\lambda_{\rm BDN}$. In order to fulfill the fourth condition in~\eqref{eq:causal1}, in our checks we have used the values $\lambda_{\rm BDN} \geq \eta$. 

In the temperature range $0\leq T\leq 200$~MeV, we have numerically found that all conditions except for one defined by Eq.~\eqref{equstability3} are satisfied. This result has been subsequently confirmed by the analytic calculations where different relations between the regulator functions have been used.  It means that the special case of the IS theory considered here is not causal.  
\section{Summary}

In this work we have found the
exact correspondence between Israel-Stewart (IS) hydrodynamic approach and first-order causal and stable (FOCS) hydrodynamics for the boost-invariant massive case with zero baryon density. The crucial assumption that allowed for this matching was that the same constant relaxation times were used in the shear and bulk sectors. Explicit expressions for the temperature dependent regulators of the FOCS theory have been given in the case where the system's equation of state is that of a relativistic massive gas. The stability and causality criteria known in the first-order approach have been applied to the Israel-Stewart framework. We have found that one of them is violated in the considered case. We note that in a very recent paper \cite{Bemfica:2020xym} the conditions for causality and stability of the IS theory have been obtained, which can be directly used in future investigations.

\medskip
{\bf Acknowledgements:} We thank J. Noronha for many illuminating comments. We would also like to thank A. Jaiswal for very important discussions. This work was supported in part by the Polish National Science Center Grants No.~2016/23/B/ST2/00717 and No.~2018/30/E/ST2/00432.
\medskip

\bibliographystyle{utphys} 


\providecommand{\href}[2]{#2}\begingroup\raggedright\endgroup

\end{document}